\documentclass[aps,prd,showpacs,eqsecnum,twocolumn,superscriptaddress,a4paper,nofootinbib]
{revtex4-1}
\usepackage{amsmath,amssymb,graphicx,color}
\usepackage{bm}
\usepackage{epsf}

\begin{document}

\title{Gravitational waves from supermassive stars collapsing to a
  supermassive black hole}

\author{Masaru Shibata} \affiliation{Center for Gravitational Physics,
  Yukawa Institute for Theoretical Physics, Kyoto University, Kyoto,
  606-8502, Japan~}

\author{Yuichiro Sekiguchi} 
\affiliation{Department of Physics, Toho
  University, Funabashi, Chiba 274-8510, Japan}

\author{Haruki Uchida}
\affiliation{Yukawa Institute for Theoretical Physics, 
Kyoto University, Kyoto, 606-8502, Japan~} 

\author{Hideyuki Umeda}
\affiliation{Department of Astronomy, Graduate School of Science,
the University of Tokyo, Tokyo, 113-0033, Japan~} 

\date{\today}
\newcommand{\beq}{\begin{equation}}
\newcommand{\eeq}{\end{equation}}
\newcommand{\beqn}{\begin{eqnarray}}
\newcommand{\eeqn}{\end{eqnarray}}
\newcommand{\pa}{\partial}
\newcommand{\vp}{\varphi}
\newcommand{\varep}{\varepsilon}
\newcommand{\ep}{\epsilon}
\newcommand{\comp}{(M/R)_\infty}
\begin{abstract}

We derive the gravitational waveform from the collapse of a rapidly
rotating supermassive star (SMS) core leading directly to a seed of a
supermassive black hole (SMBH) in axisymmetric numerical-relativity
simulations. We find that the peak strain amplitude of gravitational
waves emitted during the black-hole formation is $\approx 5 \times
10^{-21}$ at the frequency $f \approx 5$\,mHz for an event at the
cosmological redshift $z=3$, if the collapsing SMS core is in the
hydrogen-burning phase. Such gravitational waves will be detectable by
space laser interferometric detectors like eLISA with signal-to-noise
ratio $\approx 10$, if the sensitivity is as high as LISA for
$f=1$--10\,mHz. The detection of the gravitational-wave signal will
provide a potential opportunity for testing the direct-collapse
scenario for the formation of a seed of SMBHs.

\end{abstract}


\maketitle


\section{Introduction}

Clarifying the formation process of supermassive black holes (SMBHs),
which are often observed in the center of galaxies~\cite{KH2013}, is
one of the longstanding problems in astrophysics~\cite{Rees}.  One
possible scenario for this is the so-called direct-collapse scenario,
in which one supposes that a supermassive star (SMS) of mass $\agt
10^5M_\odot$ would be the progenitor for the formation of a seed of
SMBHs. Recent star-formation calculations~\cite{Hoso2013,Umeda} have
suggested that if a high mass-accretion rate with $\agt
0.1M_\odot$/yrs is preserved in the period of nuclear burning $\sim 2
\times 10^6$\,yrs, a SMS (in nuclear burning) with mass $\agt 2 \times
10^5M_\odot$ could be indeed formed.  Subsequently, the SMS core
(i.e., a central high-density isentropic region) would collapse
directly to a black hole by general-relativistic quasi-radial
instability~\cite{Iben1963,Ch1964,Fowler}. The high mass-accretion
rate, necessary for the formation of SMSs, requires primordial gas
clouds with virial temperature $\agt 10^{4}$\,K.  There are several
scenarios proposed to achieve this condition such as Lyman-Werner
radiation from nearby local star formation
region~\cite{Omukai2001,Dijkstra08} or shock heating in cold accretion
flows in the forming first
galaxies~\cite{Dekel2009,Inayoshi2012}. However, these possibilities
have not been tested by the observation yet. To test the
direct-collapse scenario, a certain observation is necessary.

The formation process of a SMBH after the collapse of a SMS is
determined by the initial condition at which the general-relativistic
quasi-radial instability sets in for the SMS.  In reality, it is
natural to consider that SMSs are rotating because they are likely to
be formed in a non-symmetric environment of a galactic center as
indicated by recent numerical simulations for the collapse of an
atomic cooling halo in the early universe
(e.g.,~Refs.~\cite{Latif2013,Regan2014,Becerra2015}). These
simulations have suggested that proto-stellar disks initially formed
in the central gas could be gravitationally unstable and fragment into
several clumps, preventing straightforward growth of mass of the
central protostar.  However, the fragments are likely to subsequently
migrate inward by gas drags and fall onto the central
protostar~\cite{Ina14,Hoso15}, leading eventually to a rotating SMS.

Motivated by this possibility, we determined the realistic conditions
for the onset of the general-relativistic quasi-radial instability of
rigidly rotating SMS cores in nuclear-burning phases~\cite{SUS16}.
Here, the reason that the SMS cores are supposed to be rigidly
rotating is that they are in nuclear-burning phases and hence in a
strongly convective phase, resulting in a uniform rotation as in
massive stars (see, e.g., Ref.~\cite{HW06}).  We find that (i) the
equation of state (EOS) for the SMS cores in hydrogen and helium
burning phases that are close to a marginally stable state against the
general-relativistic quasi-radial instability can be approximated by a
polytropic EOS with the adiabatic index $\Gamma \approx 1.335$; (ii)
the SMS cores in maximally rigid rotation are unstable against the
gravitational collapse if their mass exceeds $\approx 6.3 \times
10^5M_\odot$ in the hydrogen-burning phase and $\approx 2.3 \times
10^5M_\odot$ in the helium-burning phase; (iii) the dimensionless spin
parameter for such SMS cores is $\approx 0.8$.  These marginally
stable states, on which we focus in this paper, are plausible initial
conditions for the collapse to a seed of SMBHs.

\begin{figure*}[t]
\begin{center}
\epsfxsize=1.72in
\leavevmode
\epsffile{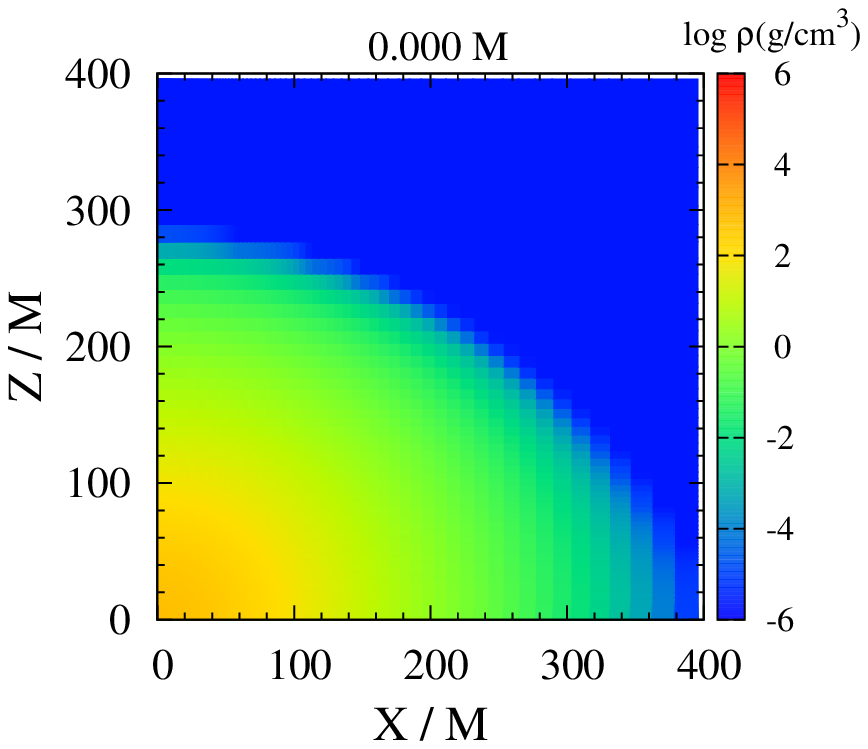}
\epsfxsize=1.72in
\leavevmode
\epsffile{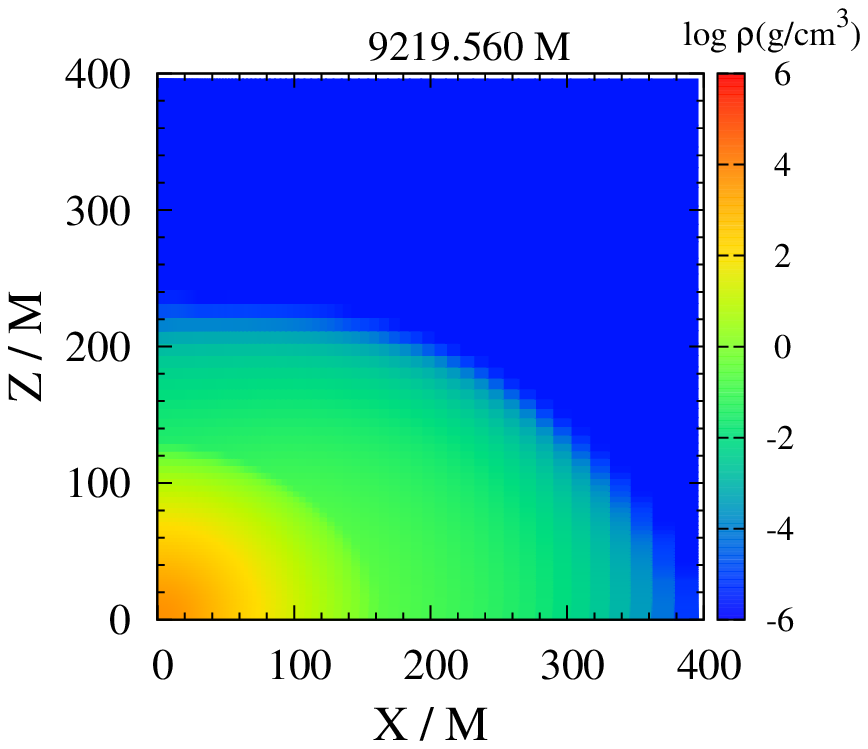}
\epsfxsize=1.72in
\leavevmode
\epsffile{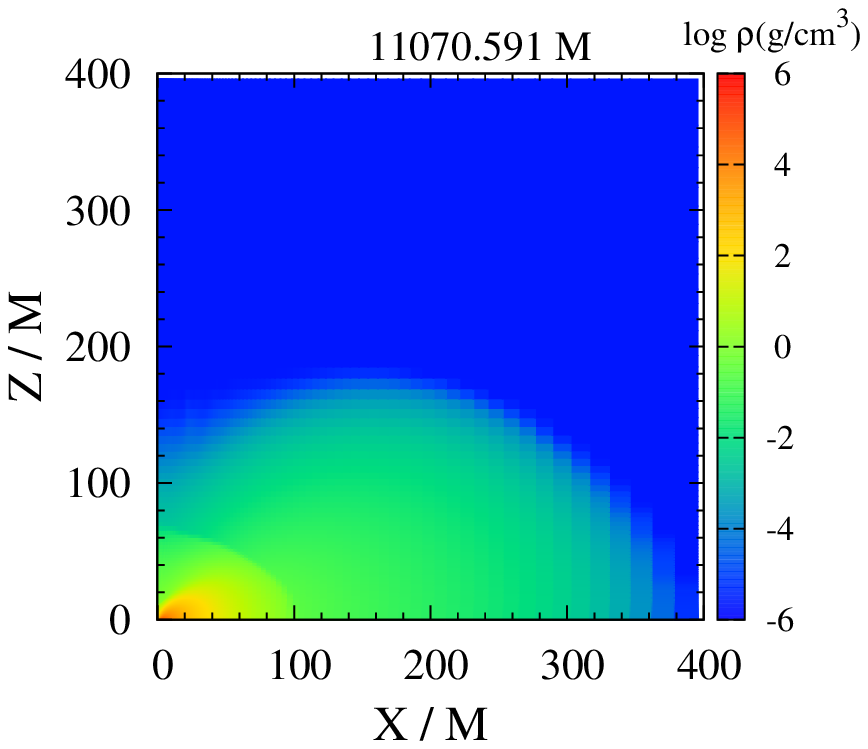}
\epsfxsize=1.72in
\leavevmode
\epsffile{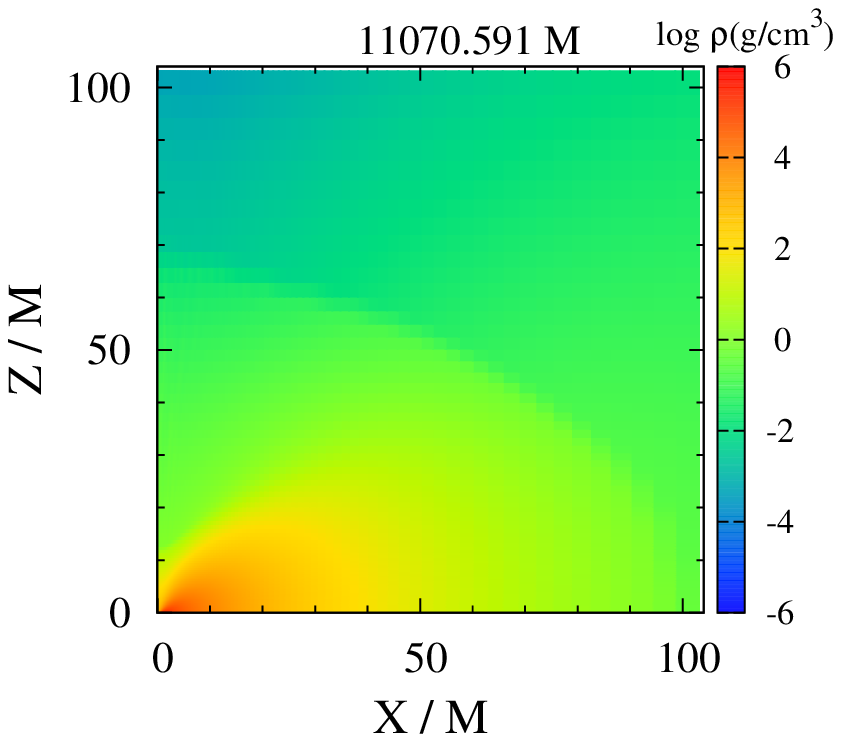} \\
\epsfxsize=1.72in
\leavevmode
\epsffile{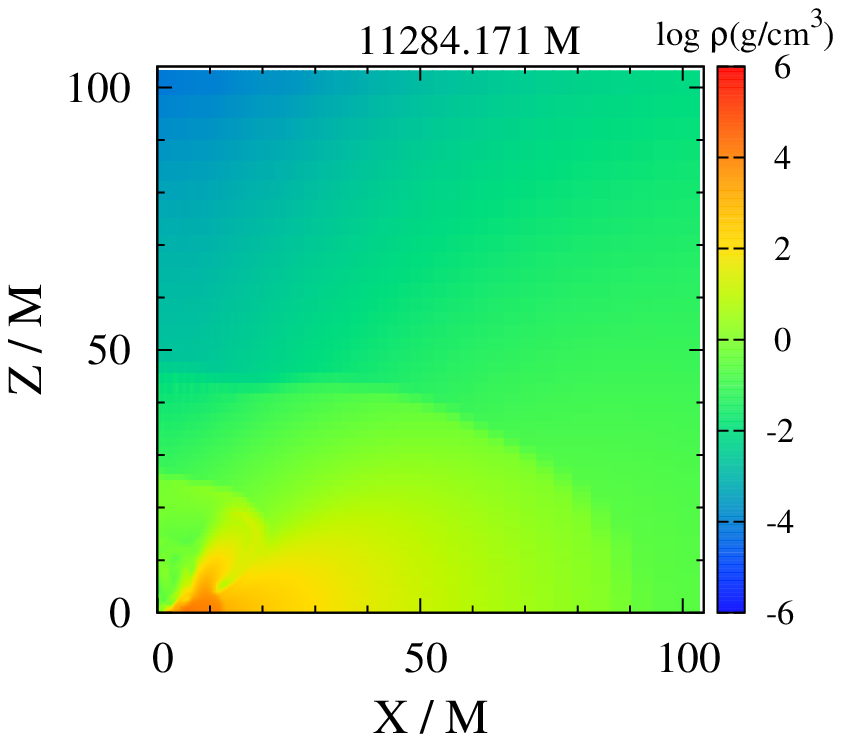}
\epsfxsize=1.72in
\leavevmode
\epsffile{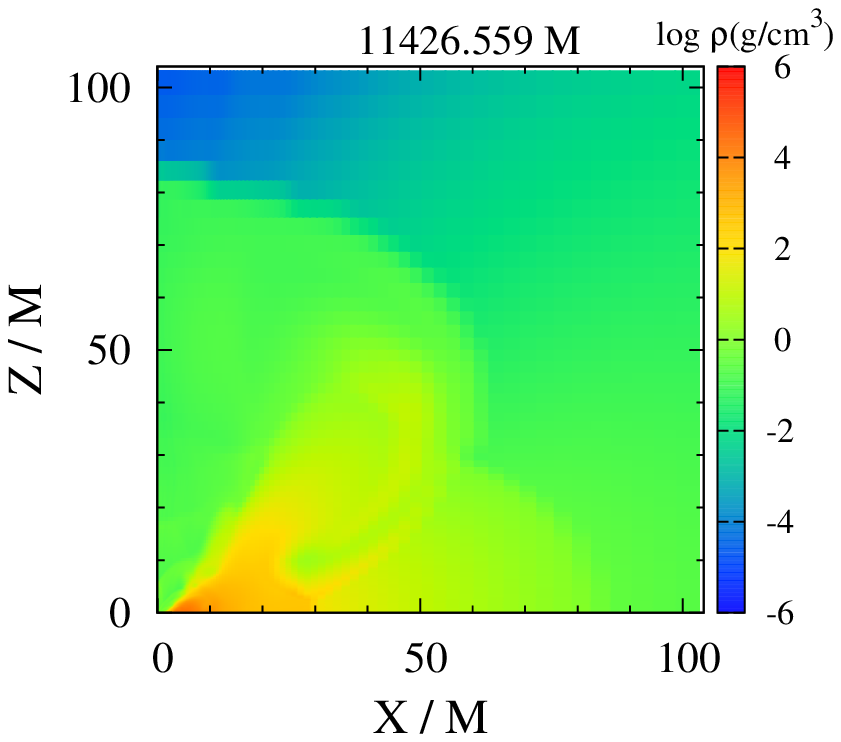}
\epsfxsize=1.72in
\leavevmode
\epsffile{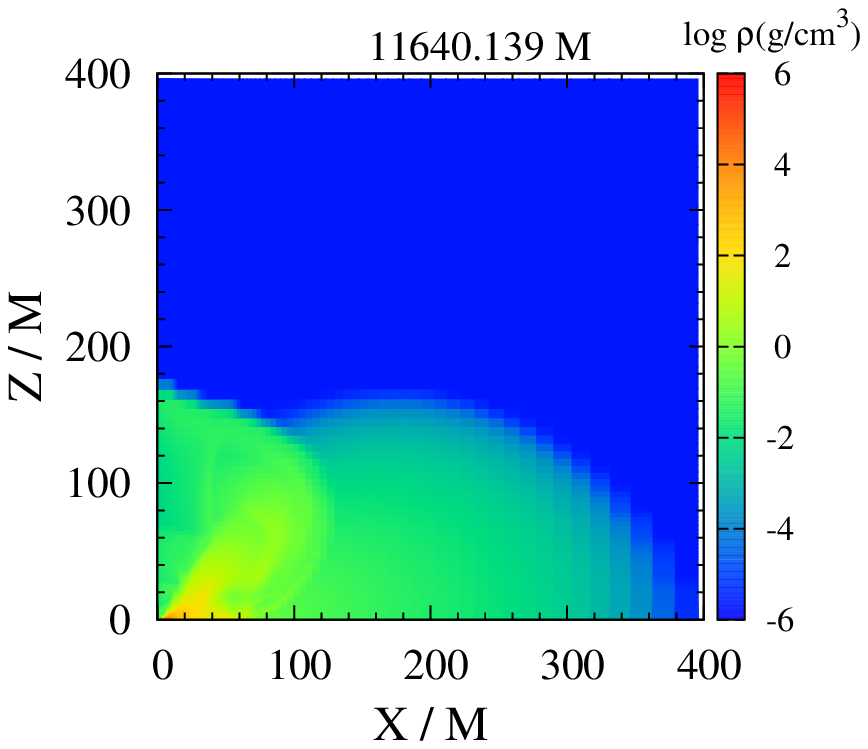}
\epsfxsize=1.72in
\leavevmode
\epsffile{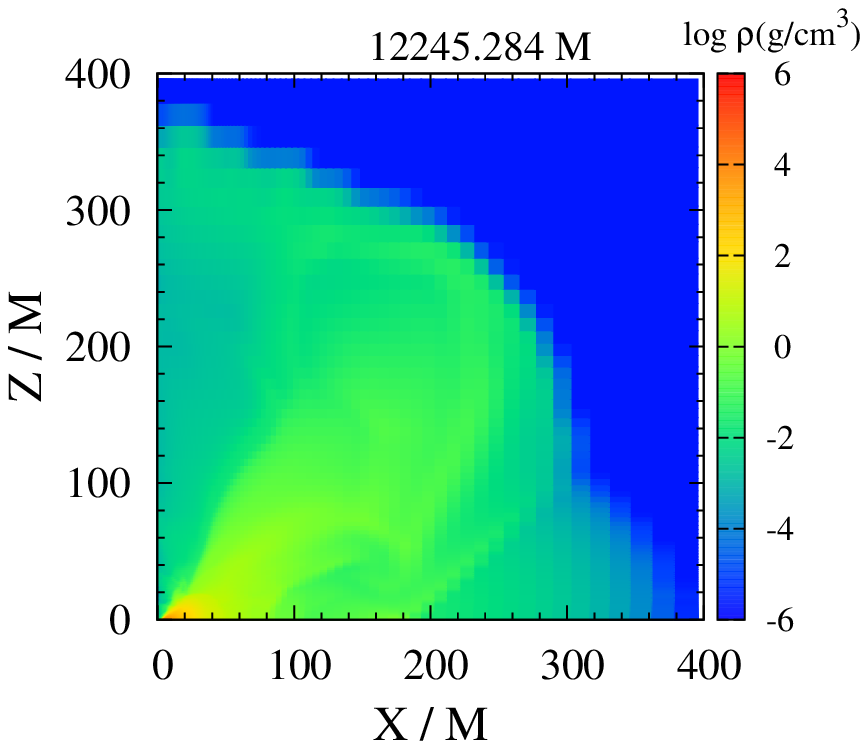}
\vspace{-4mm}
\end{center}
\caption{Snapshots of density profiles for the SMS core collapse to a
  black hole surrounded by a torus for the D2 model.  The black hole
  is formed at $t \approx 11050M$ (near the time for the 3rd and 4th
  panels) in this model. Outer boundaries of the computational domain
  are located at $533M$ along each axis. The 4th--6th panels show
  zoom-in views of the central region and the 4th panel is a zoom-in
  of the 3rd one.
\label{fig1}}
\end{figure*}

In this article, we first report a result of our new
numerical-relativity simulations for the collapse of a rapidly
rotating SMS core, focusing on gravitational waves emitted during the
black-hole formation. The direct collapse of rigidly rotating SMSs to
a SMBH has been studied by several groups~\cite{SS02,Liu07,montero}
(see also Refs.~\cite{zink,saijo2009,reis} for a different scenario)
but no group has derived the accurate gravitational waveform
associated with the black-hole formation in the scenario that we
suppose.  We show that the gravitational-wave signal is characterized
by a ringdown oscillation of the formed black hole with the frequency
$\approx 20 (M/6.3\times 10^5M_\odot)^{-1}(1+z)^{-1}$\,mHz and the
strain amplitude $\approx 5 \times 10^{-21} (M/6.3\times
10^5M_\odot)(D/25\,{\rm Gpc})^{-1}$ where $M$ is the mass of the SMS
core, $z$ is the cosmological redshift of the source, and $D$ is its
luminosity distance (which is $\approx 26$\,Gpc for $z=3$ in the
$\Lambda$CDM model). Since the best-sensitivity frequency band of
eLISA and LISA is $\approx 1$--10\,mHz~\cite{elisa,pau2012,klein15},
such a gravitational-wave signal is one of their possible targets.  We
then emphasize that the detection of this characteristic
gravitational-wave signal will be used for testing the direct-collapse
scenario for the seed-SMBH formation and that improving the
sensitivity of eLISA around 1--10\,mHz is crucial for this
purpose. Throughout this paper, we employ the units of $c=1=G$ where
$c$ and $G$ are the speed of light and gravitational constant,
respectively.

\section{Numerical results}

\begin{table}[t]
\begin{center}
\caption{Quantities for SMS core employed in this letter and for
  remnant black hole.  $\Gamma$: adiabatic index.  $M$: gravitational
  mass of the system. $\beta$: ratio of rotational kinetic energy to
  gravitational potential energy.  $J$: angular momentum. $R_e$:
  equatorial circumferential radius.  $M_{\rm BH}$ and $a_{\rm BH}$:
  mass and dimensionless spin of the black hole eventually formed.  }
\begin{tabular}{cccccccc}
\tableline\tableline
$\Gamma$ & $M~(M_\odot)$ & $\beta$ & $J/M^2$ & $R_e/M$ & $M_{\rm BH}~(M_\odot)$ 
& $a_{\rm BH}$ 
\\ \tableline
~1.335~ & ~$6.3\times10^5$~ &  0.0090 & ~0.80~ & ~423~ &
 ~$6.0\times10^5$~ & ~0.68~ 
\\ \tableline 
\end{tabular}
\end{center}
\end{table}

Our method for a solution of Einstein's equation is the same as that
in Ref.~\cite{SS2012}: We employ the original version of
Baumgarte-Shapiro-Shibata-Nakamura formulation with a puncture
gauge~\cite{BSSN}.  The gravitational field equations are solved in
the standard 4th-order finite differencing scheme.  The axial symmetry
is imposed using a 4th-order cartoon
method~\cite{cartoon,cartoon2,SS2012}, because non-axisymmetric
deformation is unlikely to be excited during the collapse in the
rigidly rotating initial condition (see, e.g., Ref.~\cite{SS05}).
Gravitational waves are extracted from the outgoing-component of the
complex Weyl scalar $\Psi_{4}$, which is expanded by a spin-weighted
spherical harmonics of weight $-2$, $_{-2}Y_{lm}(\theta,\varphi)$,
with $m=0$ in axisymmetric spacetime (see,
e.g.,~Ref.~\cite{yamamoto08}). In this work we focus only on the
quadrupole mode with $l=2$ (denoted by $\Psi_{20}$ in the following)
because it is the dominant mode.

\begin{figure}[t]
\epsfxsize=3.3in
\leavevmode
\epsffile{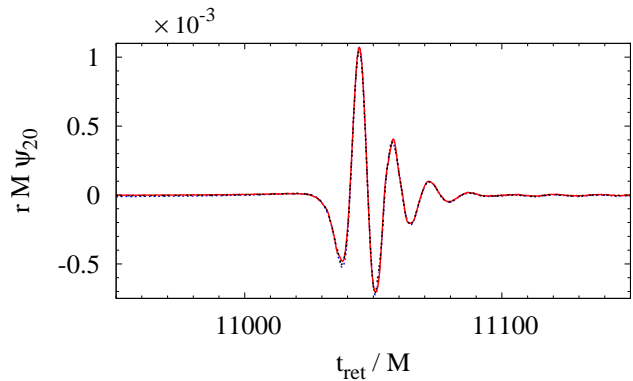}
\vspace{-3mm}
\caption{Gravitational waveforms ($l=2$ axisymmetric mode of $\Psi_4$)
  as a function of retarded time. Gravitational waves are extracted at
  $50M$ (dotted blue curve) and $75M$ (solid red curve) for the D2
  model and at $75M$ (dot-dot black curve) for the D1 model. It is
  found that these three curves agree well with each other.  Note that
  a black hole is first formed at $t_{\rm ret} \approx 11050M$ for the
  D2 model and at $t_{\rm ret} \approx 15538M$ for the D1 model, and
  hence the waveform for the D1 model is shifted by $4488M$.
\label{fig2}}
\end{figure}

A rigidly rotating SMS core near mass shedding limit is employed as
the initial condition with the polytropic EOS, $P=\kappa
\rho^{\Gamma}$, where $\kappa$, $\rho$, and $P$ are the polytropic
constant, the rest-mass density, and the pressure, and we choose
$\Gamma=1.335$ because we found it a realistic value~\cite{SUS16}:
$\Gamma$ is approximately written as $\Gamma=4/3+3.7\times
10^{-3}(M/10^5M_\odot)^{-1/2}(Y_T/1.69)$ where $M$ is the mass of the
SMS core and $Y_T$ is the total particle number per baryon, which is
1.69 for pure hydrogen plasma and 0.75 for pure helium plasma. For
rigidly rotating SMS cores that are at mass shedding limit and
marginally stable against general-relativistic gravitational collapse,
$M$ is $\approx 6.3 \times 10^5M_\odot$ and $2.3 \times 10^5M_\odot$
in the hydrogen-burning and helium-burning phases, respectively. By
appropriately setting $\kappa$, we choose $M=6.3 \times 10^5M_\odot$
in this paper (see Table I for physical quantities of the SMS
core). In this model, the central temperature is $\approx 10^{8.2}$\,K
which agrees with that in the stellar evolution
calculation~\cite{Umeda}.  The initial ratio of the polar to
equatorial axis lengths is $\approx 2/3$. During numerical evolution,
we employ the $\Gamma$-law EOS, $P=(\Gamma-1) \rho \varep$, where
$\varep$ is the specific internal energy. This is a good approximation
for the realistic EOS of the SMS core as long as we focus on the phase
up to black-hole formation because the effects of nuclear burning and
neutrino emission are minor in the standard scenario for metal-poor
SMSs~\cite{montero}.  To slightly accelerate the collapse, we
initially reduce the pressure by 1\% or 2\% uniformly (we refer to
each model as D1 and D2, respectively). We checked that for these two
depletion cases, the resulting gravitational waveforms agree well (see
Fig.~\ref{fig2}).

Numerical simulations are performed in cylindrical coordinates $(X,
Z)$, and a nonuniform grid is used for $X$ and $Z$.  Specifically, we
employ the following grid spacing (the same profile is chosen for
$Z$): for $X \leq X_{\rm in}$, $\varDelta X=\varDelta X_0=$const and
for $X > X_{\rm in}$, $\varDelta X_i=\eta \varDelta X_{i-1}$.  Here,
$\varDelta X_0$ is the grid spacing in an inner region and $X_{\rm
  in} \approx 2M$.  $\varDelta X_i:= X_{i+1}- X_{i}$ with $X_i$ the
location of $i$-th grid. At $i={\rm in}$, $\varDelta X_i=\varDelta
X_0$.  $\eta$ determines the nonuniform degree of the grid spacing.
We employ $\varDelta X_0=0.046M$ and $\eta=1.018$ for a low-resolution
run and $\varDelta X_0=0.037M$ and $\eta=1.015$ for a high-resolution
run.  We confirm that the numerical results depend only weakly on the
grid resolution. In the following, we show the results by the
high-resolution run.

Figure~\ref{fig1} displays snapshots of density profiles for the SMS
core collapse to a black hole surrounded by a torus for the D2 model.
The SMS core collapses directly to a black hole (see 1st--4th panels
of Fig.~\ref{fig1}) and 95.5\% of the total rest mass falls into the
black hole eventually irrespective of the initial pressure depletion
factor. The properties of the black hole are determined by analyzing
the area and circumferential radii of apparent horizons after the
black hole relaxes to a stationary state.  We find that the final
black-hole mass is $M_{\rm BH}\approx 6.0\times 10^5M_\odot$ and the
dimensionless spin is $a_{\rm BH} \approx 0.68$. All these values do
not depend on the initial pressure depletion factor, and also, agree
approximately with those predicted from the initial condition in the
assumption that the specific angular momentum of each fluid element is
conserved~\cite{SUS16}.  Since the SMS is rapidly rotating in this
model, $\approx 4.5\%$ of the rest mass eventually constitutes a torus
surrounding the central black hole. The geometrical thickness of the
torus is high, because shocks are formed and heat up the matter in an
inner region of the torus during its formation (see the 5th panel of
Fig.~\ref{fig1}). The inner part of the torus relaxes to a stationary
state in $1000M$ after the formation of the black hole, while the
envelope expands due to the heating by the early shock (see the
6th--8th panels of Fig.~\ref{fig1}). Exploring the subsequent
evolution of the torus and resulting signals is one of the interesting
future issues.

\begin{figure}[t]
\epsfxsize=3.2in
\leavevmode
\epsffile{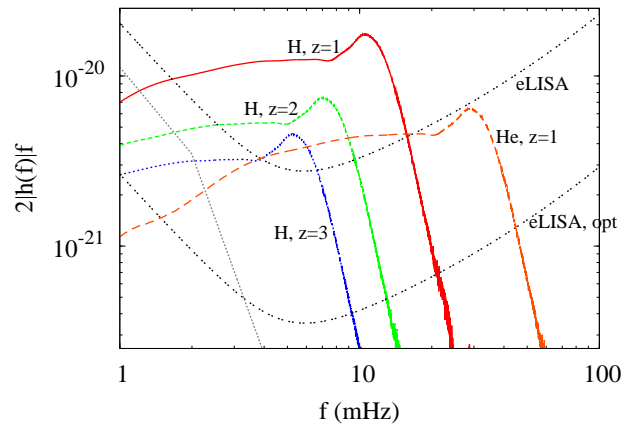}
\vspace{-3mm}
\caption{Fourier spectrum of gravitational waves: The dotted, solid,
  dashed, and long dashed curves denote the results for the
  hydrogen-burning model at $z=1$ (H, z=1), at $z=2$ (H, z=2), at
  $z=3$ (H, z=3), and helium-burning model at $z=1$ (He, z=1). We plot
  $2|h(f)|f$ because the signal-to-noise ratio (SNR) is written by
  $\int_0^{\infty} d\ln f (2|h(f)|f)^2/(S_n(f)f)$ where $S_n(f)$ is
  the one-sided noise spectrum density. The dot-dot curves denote the
  planned noise curve of eLISA (upper)~\cite{elisa} and proposed
  optimal one (lower; N2A5MxL4 of Ref.~\cite{klein15} for which the
  arm length is assumed to be 5 million km). Here, plotted is
  $\sqrt{S_n(f)f}$. The short-dotted curve denotes a model for the
  expected unresolved part of the gravitational-wave signals emitted
  by galactic binaries~\cite{klein15}. Note that the SNR for the
  hydrogen-burning cases is 5.4 and 2.2 at $z=1$ and 2 for eLISA while
  it is $\approx 43$, 17, and 10 at $z=1$, 2, and 3 for the optimally
  designed one.
\label{fig3}}
\end{figure}

Figure~\ref{fig2} displays the gravitational waveform ($\Psi_{20}$) as
a function of the retarded time, $t_{\rm ret}:=t-[r_{\rm A}+ 2M
  \ln(r_{\rm A}/2M-1)]$, where $r_{\rm A}$ is the circumferential
radius defined from the coordinate radius, $r$, by $r_{\rm
  A}:=r(1+M/2r)^2$. Note that this mode has the maximum amplitude for
the observer located on the equatorial plane and the amplitude
vanishes if the observer is located along the rotation axis. It is
found that gravitational waves are composed of a precursor associated
with long-term collapse of the SMS and of a ringdown oscillation
associated with the formed black hole. The period of the ringdown
oscillation is $\approx 16M_{\rm BH}$ and it agrees well with the
result of a linear-perturbation analysis for the black-hole
quasi-normal mode~\cite{berti2009} with $a_{\rm BH} \approx 0.68$. The
corresponding rest-frame frequency is $\approx 1/16M_{\rm BH} \approx
21 (M_{\rm BH}/6\times 10^5M_{\odot})^{-1}$\,mHz. The total energy of
gravitational waves emitted is $\Delta E \approx 1.1 \times
10^{-6}M$. Thus, the emissivity is much smaller than those in binary
black hole mergers in which $\sim 0.1M$ can be radiated (e.g.,
Ref.~\cite{lovelace}).

Figure~\ref{fig3} shows the Fourier spectrum, $|h(f)f|$, of
gravitational waves. Here, $h(f)$ is derived from 
\beqn 
h(f)=- {}_{-2}Y_{20}\int dt {2 \Psi_{20}(t) \over (2\pi f)^2} \exp(-2\pi f t)dt. 
\eeqn
Figure~\ref{fig3} is generated for the average value of $_{-2}Y_{20}
\propto \sin^2\theta$ (i.e., setting the average $\langle \sin^2\theta
\rangle=2/3$).  We choose cosmological redshifts as $z=1$, 2, and 3
for which the luminosity distance is $D\approx 6.7$, $15.8$, and
25.9\,Gpc, respectively, in the standard $\Lambda$CDM model.  We also
plot the noise curve of eLISA~\cite{elisa,pau2012} and a proposed
optimal one (N2A5MxL4 of Ref.~\cite{klein15}). This shows that the
gravitational-wave frequency associated with the ringing oscillation
of the formed black hole, $ \approx 21(1+z)^{-1}(M_{\rm BH}/6\times
10^5M_{\odot})^{-1}$\,mHz, is in the most sensitive frequency band of
these proposed space interferometers. Because the gravitational-wave
amplitude (observed along the most optimistic direction) is
approximately written as $4(M\Delta E)^{1/2}/D$, it is of the order
$10^{-20}$ for the cosmological scale with $D \sim 7$\,Gpc $(z \sim
1)$.  Since the SNR($=\int_0^{\infty} df (2|h(f)|)^2/S_n(f)$ with
$S_n(f)$ the one-sided noise spectrum density) for it is $\approx 5$
at $z=1$, the sensitivity of originally planned eLISA may not be high
enough for a confident detection of these gravitational waves, if they
are emitted for $z \agt 1$. However, if the sensitive is improved by a
factor of $\sim 10$ as discussed in Ref.~\cite{klein15} (i.e., if the
sensitive is as high as the LISA project), the SNR would be $\agt 10$
for $z \alt 3$ and a confident detection for them will be possible.
Note that SMS could be formed only in an ultra metal-poor environment,
which would be present only for high-redshift universe with $z \agt 2$
(e.g., Ref.~\cite{TFS07}). Thus, a detector as sensitive as LISA (not
eLISA) will be necessary for testing the direct-collapse scenario by
detecting gravitational waves emitted for $z \agt 2$.

For the collapse of helium-burning SMS cores, the expected core mass
is $\approx 2 \times 10^5M_{\odot}$~\cite{SUS16}. In this case, the
collapse process would be qualitatively the same as that for the
hydrogen-burning SMS core. However, the mass of the black hole formed
is about 1/3 of that in the collapse of the hydrogen-burning SMS.  As
a result, the peak frequency would be $ \approx 60(1+z)^{-1}(M_{\rm
  BH}/2\times 10^5M_{\odot})^{-1}$\,mHz with the peak strain amplitude
$\approx 1/3$ as high as that in the collapse of the hydrogen-burning
SMS core. Thus, the predicted gravitational-wave signal with $z \agt
1$ would be detectable only by an optimally-designed eLISA.  For the
collapse of oxygen-burning SMS cores, the expected highest core mass is
$\approx 2 \times 10^4M_\odot$~\cite{SUS16}, and hence, it will be
difficult to detect the gravitational-wave signal even by
optimally-designed eLISA. Since the planned sensitivity of DECIGO is
better than that of eLISA for $f \agt 20$\,mHz~\cite{YS11}, such
gravitational waves may be a possible source for DECIGO.

To summarize, by a new numerical-relativity simulation, we derive the
gravitational waveform from a rapidly rotating SMS core collapsing to
a seed of a SMBH. The predicted frequency at the Fourier-spectrum peak
is $\approx 20(1+z)^{-1}$\,mHz and the peak amplitude is $\approx 5
\times 10^{-21}$ for an event at $z=3$. This gravitational-wave signal
will be detectable by space interferometric gravitational-wave
detectors if its sensitivity is as high as LISA.  The detection of
this signal will provide a potential opportunity for testing the
direct-collapse scenario for the formation of a seed of SMBHs.


{\em Acknowledgments}: We thank N. Seto and K. Ioka for useful
discussions.  This work was supported by Grant-in-Aid for Scientific
Research (24244028, 26400220, 15H00782, 16H02183) of Japanese
MEXT/JSPS.


\end{document}